**Rheology of bacterial suspensions under confinement**

Zhengyang Liu,[1] Kechun Zhang[1] and Xiang Cheng[1,*]

1 Department of Chemical Engineering and Materials Science, University of Minnesota, Minneapolis, MN 55455, USA

* Email: xcheng@umn.edu

## Abstract

As a paradigmatic model of active fluids, bacterial suspensions show intriguing rheological responses drastically different from their counterpart colloidal suspensions. Although the flow of bulk bacterial suspensions has been extensively studied, the rheology of bacterial suspensions under confinement has not been experimentally explored. Here, using a microfluidic viscometer, we systematically measure the rheology of dilute *E. coli* suspensions under different degrees of confinement. Our study reveals a strong confinement effect: the viscosity of bacterial suspensions decreases substantially when the confinement scale is comparable or smaller than the run length of bacteria. Moreover, we also investigate the microscopic dynamics of bacterial suspensions including velocity profiles, bacterial density distributions and single bacterial dynamics in shear flows. These measurements allow us to construct a simple heuristic model based on the boundary layer of upstream swimming bacteria near confining walls, which qualitatively explains our experimental observations. Our study sheds light on the influence of the boundary layer of collective bacterial motions on the flow of confined bacterial suspensions. Our results provide a benchmark for testing different rheological models of active fluids and are useful for understanding the transport of microorganisms in confined geometries.

# Introduction

An active fluid is a suspension of self-propelled particles in fluid media with examples including a wide range of biological and physical systems ranging from swimming microorganisms (Schwarz-Linek et al. 2016), to suspensions of synthetic colloidal swimmers (Palacci et al. 2010; Palacci et al. 2013; Bricard et al. 2013) and to ATP-driven cytoskeletons (Sanchez et al. 2012; Schaller et al. 2010). With the ability of converting ambient or internal free energy to mechanical work at microscopic scales, active fluids can maintain a nonequilibrium steady state with uniform free energy input and display features drastically different from those of passive colloidal suspensions (Koch and Subramanian 2011; Marchetti et al. 2013; Bechinger et al. 2016; Elgeti, Winkler, and Gompper 2015; Saintillan and Shelley 2015). Many nonequilibrium properties of active fluids such as the emergence of collective motions (Sokolov and Aranson 2012; Wensink et al. 2012; Cisneros et al. 2011; Guo et al. 2018), giant number fluctuations (Narayan, Ramaswamy, and Menon 2007; Zhang et al. 2010) and enhanced diffusion of passive particles (Wu and Libchaber 2000; Mino et al. 2013; Morozov and Marenduzzo 2014; Peng et al. 2016; Yang et al. 2016) have been extensively studied in recent years. Among all these novel properties, the rheological response of active fluids presents arguably the most surprising phenomena, challenging our understanding of the flow of complex fluids (Saintillan 2018). By measuring the decay of large vortices and the torque on a rotating probe, Sokolov and Aranson first experimentally showed that the viscosity of bacterial suspensions in a free-standing film can reduce by a factor of seven compared to the suspending fluid without bacteria (Sokolov and Aranson 2009). Gachelin and co-workers used a Y-shaped microfluidic channel—a technique we will adopt below in our study—measured the viscosity of bulk bacterial suspensions (Gachelin et al. 2013). They showed that the viscosity of bacterial suspensions can be significantly lower than that of the suspending fluid. More recently, measurements by Lopez *et al.* using a conventional bulk rotational Couette rheometer demonstrated zero or even negative apparent viscosity in bulk bacterial suspensions at low shear rates (Lopez et al. 2015). In contrast to pusher swimmers such as bacteria, suspensions of algae, an example of puller swimmers, show a noticeable viscosity enhancement compared to suspensions of immobile algae at the same concentrations (Rafai, Jibuti, and Peyla 2010).

Several mechanisms have been proposed to explain the unusual rheology of active fluids [see (Saintillan 2018) and references therein]. The most widely circulated theory considers the coupling between the orientation of elongated bacteria modified by external shear and the intrinsic force dipoles exerted by bacteria on the suspending fluid. Orientated along the extensional quadrant of an imposed shear flow, a bacterium exerts a force dipole on the fluid, which induces a disturbance flow opposite to that induced by a passive elongated particle of the same shape. Such an effect, first explained by Hatwalne *et al.* (Hatwalne et al. 2004), leads to a reduction of the resistance of pusher suspensions to shear and the decrease of suspension viscosity. Incorporating further orientational dynamics of bacteria, continuum kinetic theories have been constructed based on the above picture, which quantitatively explained the rheology of dilute bacterial suspensions (Haines et al. 2009; Saintillan 2010; Ryan et al. 2011; Moradi and Najafi 2015; Alonso-Matilla, Ezhilan, and Saintillan 2016; Bechtel and Khair 2017). Hydrodynamic models have also been developed along a similar line (Cates et al. 2008; Giomi, Liverpool, and Marchetti 2010; Slomka and Dunkel 2017), which successfully predicted the existence of bacterial superfluids with zero apparent viscosity (Marchetti 2015). A second viscosity-reduction mechanism has recently been proposed by Takatori and Brady (Takatori and Brady 2014; Takatori, Yan, and Brady 2014; Takatori and Brady 2017). They considered the coupling between the shear flow and the swimming and rotational motion of active particles, which gives rise to an anisotropic active diffusivity in analogy of Taylor dispersion. The resulting diffusive stress stretches the fluid along the extensional direction of shear, similar to the effect of the force dipole induced by individual pusher swimmers, which leads to viscosity reduction even for spherical swimmers. Lastly, experiments on suspensions of bacteria and microtubules suggested that elongated active particles align near system boundary and form a smectic layer along confining walls in self-driven flows (Wioland, Lushi, and Goldstein 2016; Wu et al. 2017; Lushi, Wioland, and Goldstein 2014). This boundary layer collectively propels the fluid in the bulk and results in a self-sustained directional flow with zero or negative apparent viscosity.

While the rheology of bulk bacterial suspensions has been measured (Gachelin et al. 2013; Lopez et al. 2015), the effect of confinement on the rheology of bacterial suspensions has not been explored experimentally. The study of the rheology of confined bacterial suspensions is important from both fundamental and practical perspectives. First, the study shall provide crucial information for verifying different models. In particular, the boundary-layer mechanism suggests that the unusual rheology of active fluids originates from the boundary and, therefore, should strongly depend on system sizes. In comparison, both the kinetic theory and the diffusive stretching theory apply for bulk suspensions. Confinement may modify the rheology of active fluids in these theories indirectly via effects such as the change of particle orientations and density distributions near confining walls (Alonso-Matilla, Ezhilan, and Saintillan 2016). Second, various interesting collective dynamics including spontaneous directional flows (Wioland, Lushi, and Goldstein 2016; Wu et al. 2017) and stable bacterial vortices (Lushi, Wioland, and Goldstein 2014; Wioland et al. 2013; Wioland et al. 2016) have been found in confined active fluids. The consequence of these new phases on the rheology of active fluids is still unclear. Finally, confinement is frequently encountered in natural context of microbial systems, e.g. sperm cells in reproductive tracts and microorganisms in soil and biofilms (Foissner 1998; Or et al. 2007). Thus, the study on the rheology of confined bacterial suspensions will also help to understand bacterial transport in confined geometries.

Here, using *Escherichia coli* (*E. coli*) as our model active swimmers, we experimentally study the rheology of active fluids in microfluidic channels under different degrees of confinement. Our study reveals a strong confinement effect on dilute bacterial suspensions: the apparent viscosity of suspensions reduces by a factor of three when the confinement scale decreases from 60 μm down to 25 μm. We demonstrate that the effect of confinement is directly linked to the motility of bacteria. Furthermore, we also probe the microscopic dynamics of sheared bacterial suspensions such as the velocity profiles of suspension flows and the variation of bacterial density within confined channels. These microscopic measurements allow us to construct a simple model based on the boundary-layer mechanism, which qualitatively explains the experimental observations. Thus, our study provides not only systematic experimental results on the rheology of confined bacterial suspensions, but also evidence on the effect of the boundary layer on the rheology of active fluids.

## Materials and methods

### *E. coli* suspensions

In our experiments, we use a fluorescently tagged *E. coli* K-12 strain (BW25113) as our microswimmers, which carries the PKK PdnaA-GFP plasmid. These fluorescent cells allow us to image suspension flows with fluorescence and confocal microscopy. To prepare a motile *E. coli* suspension, bacteria are first cultured overnight at 37 °C in a terrific broth (TB) culture medium (tryptone 11.8 g/L, yeast extract 23.6 g/L, and glycerol 4 ml/L) supplemented with a 0.1% (v/v) selective antibiotic (ampicillin 100 mg/L). A small volume of the overnight culture is then diluted in a fresh TB culture medium (1:100) and grown at 30 °C in a shaker at 220 rpm for 6.5 hours. The culture is finally washed with a motility buffer via centrifuging (5 min, 800$g$) and set to a desired concentration of $n = 1.6 \times 10^{10}$ cells/ml. At this dilute concentration, bacteria do not show strong collective swarming (Guo et al. 2018). For an isolated wild-type

*E. coli*, it executes the so-called "run-and-tumble" motion (Berg 2004). In the "run" phase, the cell is propelled forward by a flagellar bundle at constant speed $v \approx 22$ μm/s. The straight "run" is punctuated by a sudden and rapid "tumble" at a rate $f$ on the order of 1 Hz, which randomizes the orientation of the cell. The run length of bacteria is thus given by $L = v/f$. For the wild-type bacteria we have $L = 33.1 \pm 8.1$ μm.

For control experiments, we also culture a mutant strain of *E. coli*, which exhibits only tumbling motions without run. The tumbler strain we use is RP1616 ($\Delta cheZ$), which is a derivative from RP437, a strain commonly used in chemotaxis study (Parkinson 1978). Phospho-CheY and CheZ are the two primary emzymatic complexes governing bacterial chemotactic behaviors. Phospho-CheY enhances clockwise rotation of the flagellar motors that enables the tumbling of bacteria. CheZ promotes the dephosphorylation rate of phospho-CheY to make bacteria stop tumbling and transition into the "run" phase. Thus, the function of CheZ ensures that bacterial tumbling is short and the locomotor responses to changes in chemicals are rapid. By knocking out *cheZ*, we slow down the dephosphorylation rate of phospho-CheY, which leads to the accumulation of phospho-CheY in bacteria. The excessive phospho-CheY makes bacteria keep tumbling instead of performing a run-and-tumble motion. The culturing procedure of tumblers is the same as the one for the swimmers described above. Lastly, we also test the rheology of dead bacteria as control. Bacteria are neutralized by adding 10 mM sodium azide in suspensions.

**Microfluidic viscometer**

We use a microfluidic viscometer for viscosity measurement. The same technique has been used in studying the rheology of bulk bacterial suspensions (Gachelin et al. 2013). As sketched in Fig. 1a, the viscometer consists of a symmetric Y-shape microfluidic channel with height $h$ and width $w$ in the main channel. In order to investigate the effect of confinement, we fabricate channels of different $h$ ranging from 25 μm up to 128 μm, whereas $w$ is fixed at 600 μm. The two side branches have the same height $h$ but half of the width of the main channel $w/2$. Under these conditions, the flow in the main channel satisfies the Hele-Shaw approximation (Lamb 1932), where shear gradients along the height direction ($y$) dominate the flow. We define a coordinate system in the main channel as follows: $x$ is along the flow direction, $y$ is along the height direction with dominant shear gradients, and $z$ is the vorticity direction along the width of the channel. The origin of the coordinate is set at the center of the channel with $y = [-h/2, h/2]$ and $z = [-w/2, w/2]$.

In a typical experiment, we inject a bacterial suspension of unknown viscosity and the suspending fluid of known viscosity into the microfluidic channel through the two side branches separately at the same flow rate using a syringe pump (Harvard Apparatus, 11 Elite) and two 100 μl syringes (Scientific Glass Equipment). The interface between the two fluids stabilizes downstream of the merging point of the two branches in the main channel with the width of the two fluids at $d_1$ and $d_2$, respectively. Since bacteria constantly migrate across the interface from the suspension into the suspending fluid, the interface gradually smooths out along the flow. We measure $d_1$ and $d_2$ at the position where the interface is stable and sharp, typically 500 to 1000 μm downstream of the merging point (Fig. 1b). It can be shown that the viscosity ratio of the two miscible fluids in the channel, $\eta_1/\eta_2$, is equal to the width ratio $d_1/d_2$ (Guillot et al. 2006; Nghe, Tabeling, and Ajdari 2010) (see also Discussions below). We test the accuracy of the viscometer at different $h$ by measuring the viscosity of water-glycerol mixtures. The results from the microfluidic viscometer quantitatively agree with the known viscosity of the mixtures and are independent of the channel height in the range of our experiments (Figs. 1c and d).

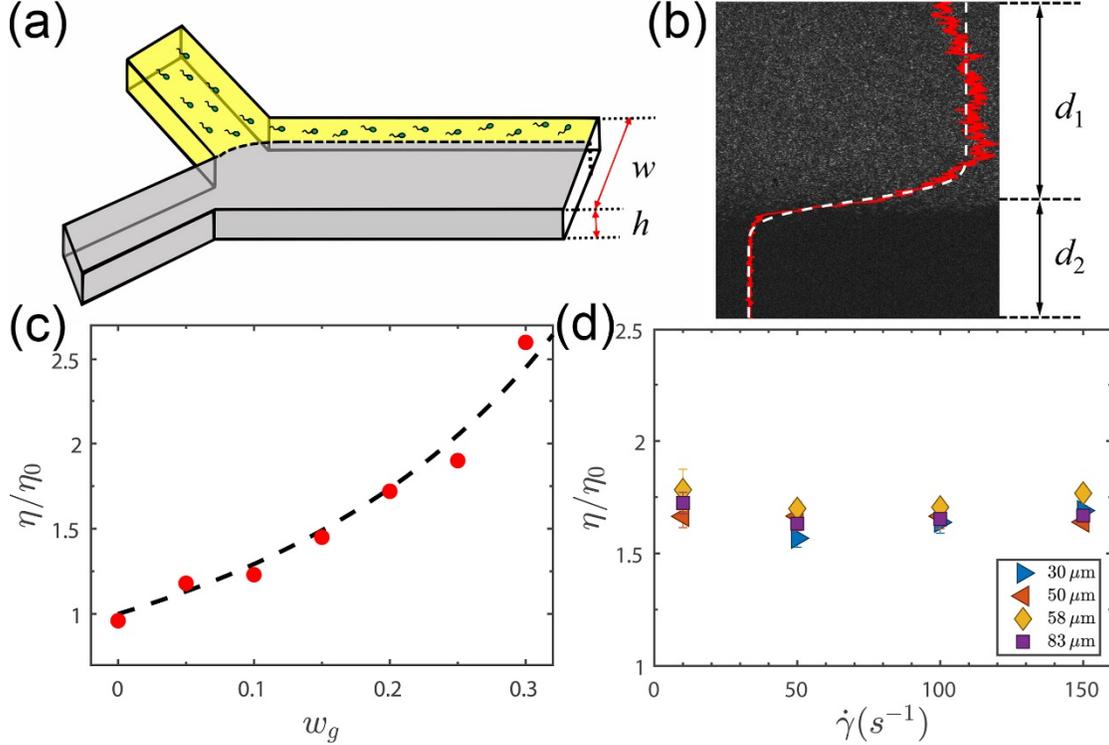

**Fig 1** Microfluidic viscometer. (a) Schematic of the Y-shaped microfluidic channel. (b) Image of two fluids in the channel. The upper region is a bacterial suspension with width $d_1$, whereas the lower region is the suspending fluid with width $d_2$. The red line shows the intensity profile across the channel. The white dashed line is the error function fit. (c) Viscosity of water-glycerol mixtures at different glycerol weight fractions, $w_g$. Shear rate $\dot{\gamma} = 100$ s$^{-1}$. The dashed line is the literature value. (d) Viscosity of a water-glycerol mixture (20 w% glycerol) as a function of shear rates at different channel heights $h$. The literature value of the viscosity of the mixture is 1.74 mPa·s at 20°C. $h$ is indicated in the figure.

In this study, we take the nominal wall shear rate $\dot{\gamma} \equiv 6Q/(d_1 h^2)$ as a characteristic shear rate on bacterial suspensions, where $Q$ is the control flow rate. The formula is exact when the velocity profiles of suspensions are parabolic following the Hagen-Poiseuille law. The average of the magnitude of shear rates in the channel is then $\dot{\gamma}/2$. For non-parabolic profiles, the formula can be simply treated as a definition of the characteristic shear rate in the channel. For each channel height $h$, we decrease $Q$ from 100 μl/h to 1 μl/h to examine the rheological response of bacterial suspensions at different $\dot{\gamma}$. At an even lower $Q$, the interface between the suspensions and the suspending fluid becomes unstable and displays non-planar longitudinal variations, which prevents us from measuring the width ratio of the two fluids accurately. Such an instability may indicate nonzero normal stress differences in bacterial suspensions (Hinch et al. 1992; Brady and Carpen 2002; Saintillan 2010).

**Image acquisition and analysis**

Florescence microscopy is used to take movies of the microfluidic flows at the center of the channel at $y = 0$. Movies are acquired at 30 frames per second (fps) with a sCMOS camera (Andor, Zyla) through an inverted microscope (Nikon, Ti-E) using a 10× objective. Raw images are first processed by a variance filter to enhance the contrast between bacterial suspensions and the suspending fluid (Fig. 1b). For each image, we calculate the sum of the pixel intensity in each row and then obtain an intensity profile of the image along the width of the channel (*z*) (Fig. 1b, red curve). By fitting the intensity profile with an error function (Fig. 1b, white curve), we identify the position of the interface between the two fluids as the reflection point of the error function. This image analysis routine is implemented using a custom MATLAB program.

To obtain the flow profiles of suspensions at different $h$ and $\dot{\gamma}$, we add fluorescent polystyrene (PS) colloids of 1 μm in diameter into bacterial suspensions. At 0.001 v% in the final mixture, the concentration of PS particles is so low that the presence of the particles should not affect the flow of bacterial suspensions. We use fast confocal microscopy to measure suspension flows at different heights above the bottom wall of microfluidic channels away from the side walls. At each height, a movie of 10–20 s is taken at 100 fps using a 60× objective. The velocity of fluid flows at a certain *y*, $V(y)$, is extracted by tracking the motion of colloids in the imaging plane using Particle Tracking Velocimetry (PTV). In addition, we also measure the average velocity of bacteria, $V_{bac}(y)$, via Particle Imaging Velocimetry (PIV), where bacteria, instead of colloids, are used as tracer particles. Both the velocity profile of fluid flows and the velocity profile of bacteria along *y* can then be compiled from a series of measurements at different heights. The measurements on $V(y)$ and $V_{bac}(y)$ yield not only the flow profile of suspensions but also the relative motions between bacteria and the suspending fluid. The disturbance flow $V_d(y)$ defined below can be obtained as $V_d(y) = V(y) - V_{bac}(y)$.

The number of bacteria at each height can also be estimated from these movies through direct counting, which allows us to calculate 2D bacterial density *n*.

## Results

### Confinement effect

Figure 2a shows the relative viscosity of bacterial suspensions, $\eta/\eta_0$, as a function of shear rates for channels of different heights. Here, $\eta$ is the viscosity of bacterial suspensions and $\eta_0$ is the viscosity of the suspending fluid. For channels with $h \gtrsim 60$ μm, the flow curves of different heights collapse into a master curve, giving the rheological response of bulk bacterial suspensions. At low shear rates, suspensions show strong shear thickening. When $\dot{\gamma} \lesssim 10$ s$^{-1}$, the viscosity of bacterial suspensions is below that of the suspending fluid, a defining feature of the rheology of pusher suspensions. At the lowest shear rate of our experiments $\dot{\gamma} = 1$ s$^{-1}$, the viscosity is about half of the viscosity of the suspending fluid, quantitatively agreeing with previous experiments (Gachelin et al. 2013). However, for channels with $h < 60$ μm, the flow curves separate from each other at low shear rates, indicating a strong confinement effect at small *h*. At high shear rates, the relative viscosity at different heights approaches to a plateau independent of *h*. This absence of the confinement effect at high shear rates suggests that the effect is linked to bacterial motility. In the limit of high shear rates, the active stress arising from the hydrodynamic stresslet and the diffusive stress is negligible compared to the passive stress induced by the rigid elongated body of *E. coli* on the fluid (Takatori and Brady 2017; Saintillan 2018). Thus, the viscosity of active suspensions in the limit of high shear rates should be the same as that of suspensions of passive elongated particles of the same shape, which does not show obvious confinement effect. We also observe weak shear thinning in this regime, presumably arising from the shear-induced alignment of elongated particles (Egres, Nettesheim, and Wagner 2006). The degree of shear thinning, however, is weaker than that reported in the previous study (Gachelin et al. 2013).

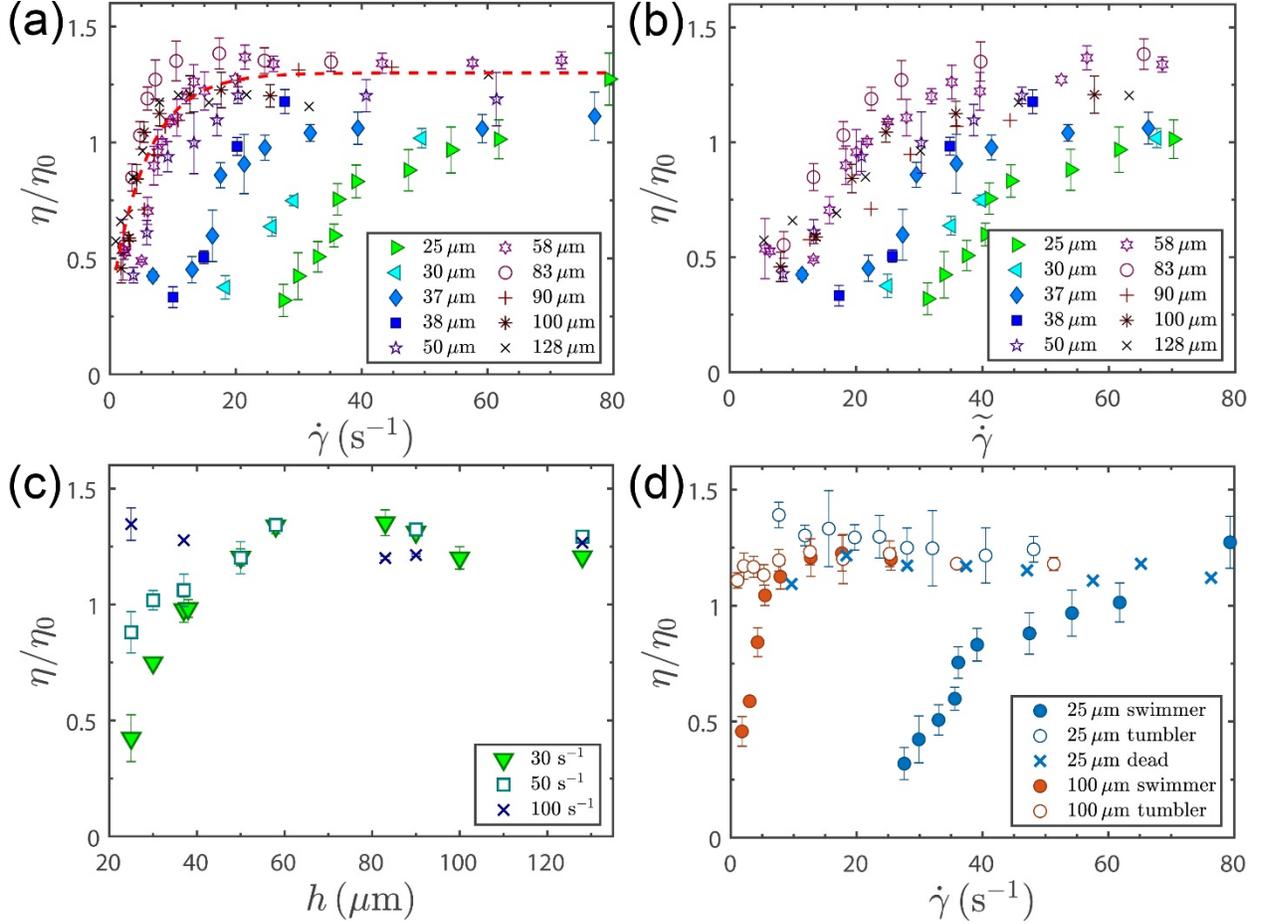

**Fig. 2** Relative viscosity of bacterial suspensions, $\eta/\eta_0$. (a) $\eta/\eta_0$ as a function of shear rates, $\dot{\gamma}$, in channels of different heights, $h$. $h$ is indicated in the figure. The dashed line is a fitting for the bulk samples with $\eta/\eta_0 = 1.3 - \exp(-0.24\dot{\gamma})$. (b) $\eta/\eta_0$ as a function of dimensionless shear rates, $\tilde{\dot{\gamma}}$, in channels of different $h$. $\tilde{\dot{\gamma}} \equiv \dot{\gamma}h/v$, where bacterial swimming speed $v = 22$ μm/s. (c) $\eta/\eta_0$ as a function of $h$ at different $\dot{\gamma}$, which are indicated in the figure. (d) $\eta/\eta_0$ of suspensions of active swimmers, tumblers and dead bacteria. Two different heights $h = 25$ μm and 100 μm are used.

    We have attempted to rescale the relative viscosity by normalizing the shear rate by the characteristic run-time of bacteria, which determines the diffusive stress of active particles (Takatori and Brady 2017). The dimensionless shear rate $\tilde{\dot{\gamma}}$ is defined as $\tilde{\dot{\gamma}} \equiv \dot{\gamma}/f$, where $f$ is the tumbling frequency of bacteria. For bulk samples with $h > 60$ μm, $f \approx 0.5$ Hz is an intrinsic property of bacteria and therefore a constant. Since the relative viscosity has already collapsed together when plotting against $\dot{\gamma}$, adding a constant prefactor should not change the quality of data collapsing. For confined samples, the tumbling frequency of bacteria is determined by the system size. In a simple picture, $f$ should be replaced by $v/h$. Although different data sets show better collapse when plotted against $\tilde{\dot{\gamma}}$, they are still well separated (Fig. 2b). The result suggests that other factor(s) in addition to the reorientation of bacteria need to be considered in order to fully interpret our experiments on strongly confined samples.

The confinement effect is even better illustrated in Fig. 2c, where the relative viscosity of bacterial suspensions as a function of channel heights is directly plotted at three different shear rates. At the lowest shear rate, the viscosity increases by a factor of three when the confinement length increases from 25 μm to 60 μm. Above $h \approx 60$ μm comparable to the run length of bacteria, the viscosity plateaus and becomes independent of $h$. At the moderate shear rate, the increasing trend is less pronounced. At the highest shear rate, the height dependence completely vanishes.

To further demonstrate that the confinement effect arises from bacterial motility, we also conduct control experiments comparing the viscosity of suspensions of active swimmers, tumblers and immobile bacteria (see Materials and methods). The viscosity of the three types of suspensions are examined in both bulk and confined systems. Figure 2d shows that the viscosity of active swimmers exhibits strong shear thickening in both bulk and confined channels, whereas the viscosity of tumblers and immobile bacteria weakly depends on shear rates. As expected, the viscosity reduction originates from the motility of bacteria. More importantly, the confinement effect disappears for suspensions of tumblers and immobile bacteria. The flow curves at $h = 100$ μm and 25 μm are quantitatively the same within experimental errors. Hence, we confirm that the motility of bacteria is the direct cause of the confinement effect.

**Microscopic dynamics**

Hydrodynamic interactions between bacteria and external shear flows can profoundly modify the swimming behaviors of bacteria, leading to interesting phenomena such as rheotaxis (Marcos et al. 2012), heterogeneous bacterial distributions (Rusconi, Guasto, and Stocker 2014) and upstream swimming along confining walls (Hill et al. 2007; Nash et al. 2010; Costanzo et al. 2012; Kaya and Koser 2012). These microscopic bacterial dynamics may strongly affect the macroscopic rheology of bacterial suspensions. Hence, we also investigate the microdynamics of bacterial suspensions at microscopic scales in microfluidic channels of different heights.

**Velocity profiles**

First, we measure the velocity profiles of bacterial suspensions along $y$ in microfluidic channels of different heights at low shear rates (Fig. 3), where the confinement effect is most pronounced. Near the center of the channels, the velocity profiles of fluid flows, $V(y)$, are parabolic, consistent with the Hagen-Poiseuille law. However, near the bottom and top walls, the velocity of fluid flows measured from passive colloidal tracers noticeably deviates from the parabolic profile and is significantly higher than the velocity of bacteria, $V_{bac}(y)$ (see Materials and methods). Thus, there exist a boundary layer near the confining walls, where bacteria swim against the background flow and exhibit strong relative motions. The thickness of these boundary layers, where strong relative motions of bacteria can be observed, is about 5 μm, independent of $h$. Thus, as $h$ decreases, the boundary layer plays a more important role in suspension flows. Interestingly, the velocity profiles of bacteria, $V_{bac}(y)$, remain parabolic for different $h$, satisfying the no-slip condition at the walls. Such a feature is crucial for *E. coli* in natural environments, where they need maintain their locations in the lower intestine of their hosts against expelling flows.

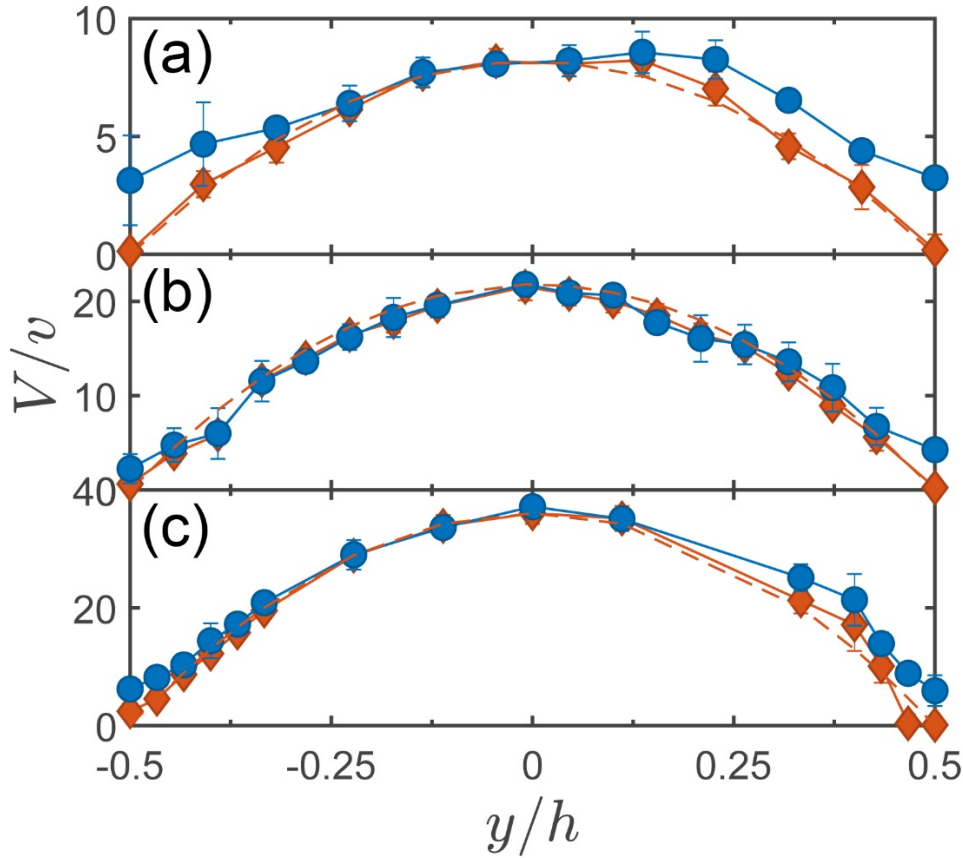

**Fig. 3** Velocity profiles of fluid flows and bacteria at different channel heights $h$. The blue disks are the velocity profiles of fluid flows measured from passive colloidal tracers. The red rhombuses are the velocity profiles of bacteria measured from bacteria. (a) $h = 30$ μm and $\dot{\gamma} = 25$ s$^{-1}$, (b) $h = 50$ μm and $\dot{\gamma} = 36$ s$^{-1}$, (c) $h = 83$ μm and $\dot{\gamma} = 38$ s$^{-1}$. Dashed lines are the fittings of parabolic profiles. Velocity is normalized by bacterial velocity $v = 22$ μm/s.

**Upstream swimming bacteria**

The difference in the velocity profiles of fluid flows and bacteria indicates the existence of boundary layers near the confining walls consisting of upstream swimming bacteria against imposed shear flows, an observation in agreement with previous experiments (Hill et al. 2007; Kaya and Koser 2012) and simulations (Costanzo et al. 2012; Chilukuri, Collins, and Underhill 2014; Ezhilan and Saintillan 2015; Nash et al. 2010). To illustrate the phenomenon directly, Figure 4a shows the trajectories of colloidal and bacterial tracers near the bottom wall. The relative motions between bacteria and the background fluid can be clearly identified (see also Supplementary Movie). The orientation of bacteria against the bottom and top walls can be estimated from the images showing bacteria near the side walls (Fig. 4b). We plot the distribution of the orientation angle of bacteria against the side walls (Fig. 4c), which strongly biases toward acute angles with the mean at 20°.

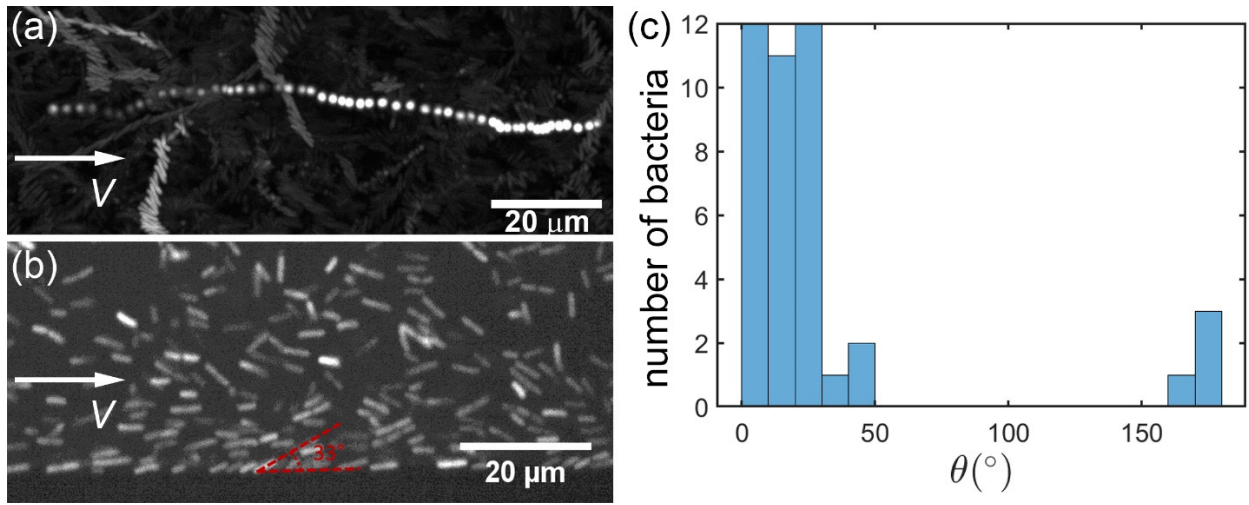

**Fig. 4** Upstream swimming bacteria. (a) Trajectories of a passive colloidal tracer and bacteria (see also Supplementary Video). The maximal intensity of 54 frames over a time interval of 17.5 s is projected onto a single image to show the trajectories. While the spherical passive tracer is transported by the fluid flow, many bacteria swim against the flow and exhibit only transverse motions. The channel height $h = 30$ μm. The shear rate $\dot{\gamma} = 24$ s$^{-1}$. (b) Bacteria near the side wall, locating at the bottom of the image. The channel height $h = 50$ μm. The shear rate $\dot{\gamma} = 7.8$ s$^{-1}$. The direction of the flow is indicated in the images. The orientation of a single bacterium at $\theta = 33°$ is indicated. (c) Distribution of the orientation of bacteria next to the side wall. The total number of bacteria counted is 42.

**Bacterial density distributions**

Finally, we also measure the number density of bacteria in microfluidic channels. Figure 5a shows the density distribution of bacterial suspensions along $y$ at a low shear rate. In consistency with previous experiments (Hill et al. 2007; Berke et al. 2008; Li and Tang 2009), we find that bacteria accumulate near the confining walls, an effect arising from the coupling between self-propulsion and steric interactions (Ezhilan and Saintillan 2015). Hydrodynamic interactions between bacteria and solid surfaces also enhance the accumulation (Berke et al. 2008). Furthermore, Figure 5b shows bacterial concentrations at the bottom confining wall as a function of shear rates. The concentration decreases with increasing shear rates, leading to more uniform density profiles at high shear rates (Ezhilan and Saintillan 2015).

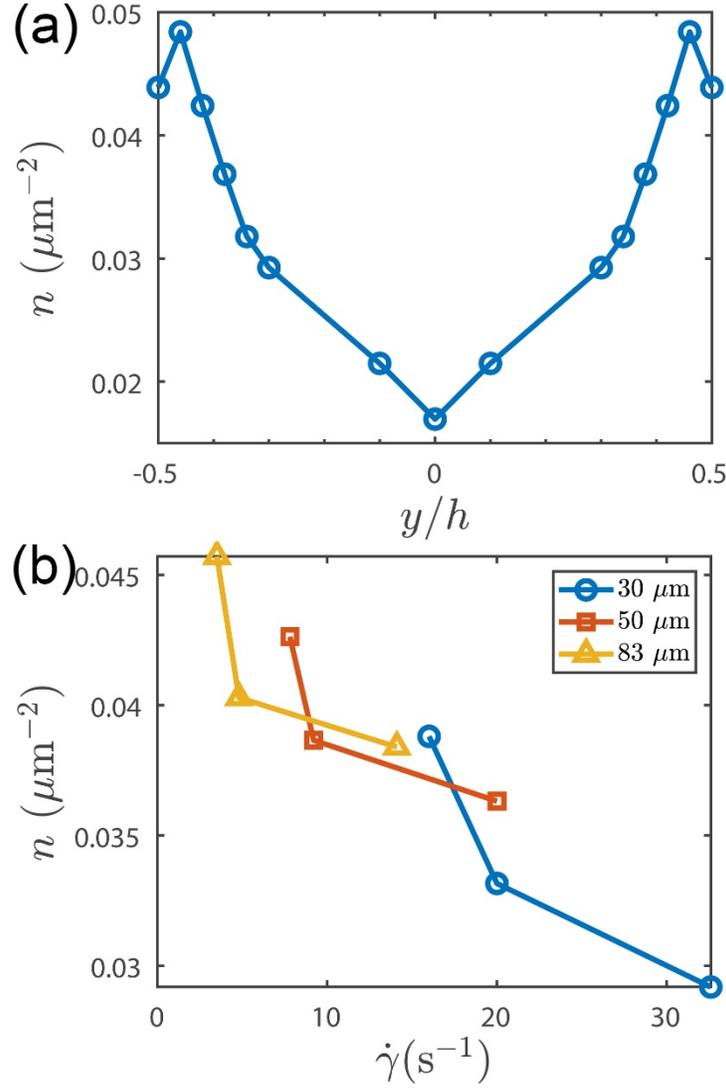

**Fig. 5** Bacterial density distribution in microfluidic channels. (a) Bacterial density across a channel of height $h = 50$ μm. The shear rate is $\dot{\gamma} = 7.8$ s$^{-1}$. (b) Bacterial density at the bottom wall as a function of $\dot{\gamma}$. Different symbols are from channels of different heights, $h$, which are indicated in the figure.

## Discussions

### Data analysis

We perform a simple data analysis to better illustrate the origin of the unusual rheology of bacterial suspensions under confinement. First, we shall recapitulate the calculation for the viscosity ratio of the two miscible fluids in a bulk Y-shaped microfluidic channel. Within the Hele-Shaw approximation, the average velocities of Fluid 1 and Fluid 2 over the channel height $y$ are given by (Lamb 1932)

$$\langle V_1 \rangle = -\frac{1}{12}\frac{h^2}{\eta_1}\frac{\partial P}{\partial x} \quad \text{and} \quad \langle V_2 \rangle = -\frac{1}{12}\frac{h^2}{\eta_2}\frac{\partial P}{\partial x}. \tag{1}$$

where $\partial P/\partial x$ is the pressure gradient driving the flow, which is the same in the two branches. $\eta_1$ and $\eta_2$ are the viscosity of Fluid 1 and Fluid 2, respectively. Thus,

$$\langle V_1 \rangle \eta_1 = \langle V_2 \rangle \eta_2. \tag{2}$$

In addition, the flow rates of the two fluids are the same,

$$\langle V_1 \rangle d_1 h = \langle V_2 \rangle d_2 h. \tag{3}$$

Combining Eq. (2) and (3), we obtain the desired relation,

$$\frac{\eta_1}{\eta_2} = \frac{d_1}{d_2}.$$

In our experiments, Fluid 1 is a bacterial suspension with an unknown viscosity of $\eta$. As the reference fluid, Fluid 2 is the suspending fluid with a known viscosity $\eta_0$. The relative viscosity of the bacterial suspension is thus (Gachelin et al. 2013)

$$\frac{\eta}{\eta_0} = \frac{d_1}{d_2}, \tag{4}$$

where $d_1$ is the width of the bacterial suspension and $d_2$ is the width of the suspending fluid.

For a confined bacterial suspension, there should be an extra contribution to the fluid flow arising from the coupling between bacterial motility and the confining surfaces of the channel. We assume this extra disturbance flow linearly superposes to the pressure-driven Poiseuille flow at low Reynolds numbers:

$$\langle V \rangle = -\frac{1}{12}\frac{h^2}{\eta_b}\frac{\partial P}{\partial x} + \langle V_d \rangle. \tag{5}$$

Here, $\langle V_d \rangle$ is the average strength of the boundary-driven disturbance flow. $\eta_b$ is the viscosity of bulk bacterial suspensions without the influence of system boundary. Numerically, we obtain $\eta_b$ by fitting our experiments on bulk samples with $h > 60$ μm using an exponential function, which gives $\eta_b/\eta_0 = 1.3 - \exp(-0.24\dot\gamma)$ (Fig. 2a). In bulk samples, the disturbance flow from the boundary is negligible with $\langle V_d \rangle \approx 0$. The non-trivial rheology results from the active hydrodynamic stresslets and diffusive stresses (Alonso-Matilla et al 2016; Takatori and Brady 2017). We shall focus on the confined systems in our analysis below, where boundary-driven disturbance flows dominate. Note that since the characteristic shear rate at which shear thickening occurs in bulk samples is $1/0.24 \approx 4$ s$^{-1}$ smaller than the lowest shear rate we can achieve in confined channels when $h < 60$ μm (Fig. 2a), the shear thickening effect of bulk samples should not strongly affect our analysis of $\langle V_d \rangle$ below. Quantitatively similar results on $\langle V_d \rangle$ are indeed obtained if we fix $\eta_b/\eta_0 = 1.3$ in Eq. (5), i.e., the high-shear-rate plateau of the relative viscosity.

With the assumption of Eq. (5) as well as the average velocity of the suspending fluid, which is Newtonian following

$$\langle V_0 \rangle = -\frac{1}{12}\frac{h^2}{\eta_0}\frac{\partial P}{\partial x}, \tag{6}$$

we have

$$\langle V \rangle = \frac{\eta_0}{\eta_b}\langle V_0 \rangle + \langle V_d \rangle. \tag{7}$$

Furthermore, the flow rates in the two fluids should be the same as before,

$$\langle V \rangle d_1 h = \langle V_0 \rangle d_2 h,$$

which leads to

$$\langle V \rangle = \frac{\eta_0}{\eta_b} \frac{d_1}{d_2} \langle V \rangle + \langle V_d \rangle.$$

Experimentally, the relative viscosity, $\eta/\eta_0$, is measured via the width ratio of the two fluids. Hence, by definition

$$\frac{\eta}{\eta_0} \equiv \frac{d_1}{d_2} = \frac{\eta_b}{\eta_0}\left[1 - \frac{\langle V_d \rangle}{\langle V \rangle}\right]. \qquad (8)$$

Finally, applying the definition of the characteristic shear rate,

$$\dot{\gamma} \equiv \frac{6Q}{d_1 h^2} = \frac{6\langle V \rangle}{h},$$

we have

$$\frac{\eta}{\eta_0} = \frac{\eta_b}{\eta_0}\left[1 - \frac{6\langle V_d \rangle}{h\dot{\gamma}}\right]. \qquad (9)$$

We fit our experimental results using the above equation, where $\eta_b/\eta_0$ is given by the exponential function for bulk samples as discussed above. $\langle V_d \rangle$ is taken as a fitting parameter, which we assume is independent of $\dot{\gamma}$. We find that $\langle V_d \rangle$ decreases with increasing $h$ and shows an approximate power-law scaling, $\langle V_d \rangle \sim h^{-\alpha}$, with $\alpha = -1.9 \pm 0.6$ (Fig. 6a). $\langle V_d \rangle$ obtained from the fitting quantitatively agrees with direct measurements based on the velocity profiles (Fig. 3), where the disturbance flow can be extracted by subtracting the parabolic bacterial flow from the total fluid flow, $V_d(y) = V(y) - V_{bac}(y)$. Using $\langle V_d \rangle$, a good collapse of data can be achieved by plotting the viscosity ratio $\eta/\eta_b$ versus the inverse dimensionless shear rate $\langle V_d \rangle/h\dot{\gamma}$ (Fig. 6b). A linear trend predicted by Eq. (9) can be clearly identified. We shall discuss the possible origin of $\langle V_d \rangle$ below.

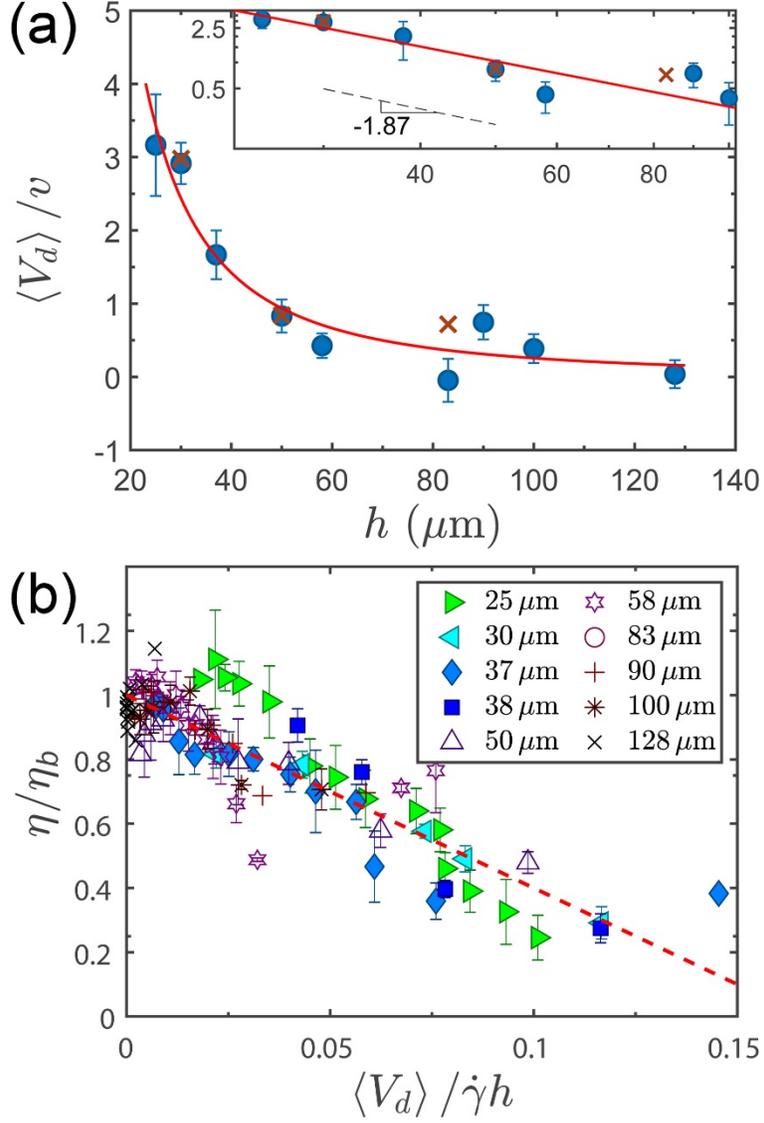

**Fig. 6** Scaling of relative viscosity. (a) The average velocity of the boundary-driven disturbance flows, $\langle V_d \rangle$, as a function of the channel height, $h$. Blue disks are from the fitting of Eq. (9). Red crosses are from the direct velocity-profile measurements (Fig. 3). $\langle V_d \rangle$ is normalized by the swimming speed of bacteria $v = 22$ μm/s. The solid red line is a power-law fitting, $\langle V_d \rangle \sim h^{-1.87}$. Inset shows the same data in a log-log plot. (b) Viscosity ratio, $\eta/\eta_b$, as a function of inverse dimensionless shear rates, $\langle V_d \rangle/(\dot{\gamma} h)$. $h$ is indicated in the plot. The dashed line indicates $y = 1 - 6x$, the prediction of Eq. (9).

**Effect of active hydrodynamic and diffusive stresses**

Using a kinetic theory with active hydrodynamic stresslets, Alonso-Matilla and co-workers studied the effect of confinement on the rheology of bacterial suspensions (Alonso-Matilla, Ezhilan, and Saintillan 2016). Since bacterial density is higher near the confining walls (Fig. 5a), as the system size reduces, the regions with higher bacterial density near the walls start to overlap near the center of microfluidic channels. As a result, relatively fewer bacteria are subject to the high shear near the walls, mitigating the effect of shear alignment and viscosity reduction. Hence, the viscosity of confined bacterial suspensions was predicted to be higher than that of bulk suspensions at the same shear rate, which is opposite to our experimental observations.

The effect of diffusive stresses in confined systems is more complicated. The hydrodynamic stress from active stresslets scales as $\sigma^H \sim n\zeta va$, which therefore does not directly depend on the confinement except via bacterial density $n$ as discussed above. Here, $\zeta$ is the drag coefficient and $a$ is the characteristic size of bacteria. In comparison, the diffusive stress scales as $\sigma^S \sim n\zeta vL$, where $L$ is the run length of bacteria (Takatori and Brady, 2017). In confined systems, bacterial run length is determined by the size of system. In a simple picture, $L$ should be replaced by the size of system $h$ as $\sigma^S \sim n\zeta vh$, which gives rise to a direct system-size dependence. The magnitude of $\sigma^S$ decreases with increasing confinement. Whether the viscosity of bacterial suspensions increases or decreases with $h$ depends on the sign of $\sigma^S$. In the low Peclet number (Pe) limit where the diffusive stress dominates, $\sigma^S$ is negative. As a result, when $h$ decreases, the magnitude of $\sigma^S$ decreases and the viscosity of suspensions increases with decreasing $h$, opposite to our experiments. However, for intermediate Pe that are more relevant to our experiments, $\sigma^S$ becomes positive (see Fig. 2 of Takatori and Brady 2017), which leads to the reduction of viscosity in confined systems, consistent with our experiments. Nevertheless, for suspensions of spherical particles, where the analytical solution of the viscosity is available (Eq. 5 of Takatori and Brady 2017), the decrease of viscosity is less than a factor of 1.1 when we decrease $h$ from 50 μm to 25 μm even at the optimal shear rates of 2.1 s$^{-1}$ that leads to the largest viscosity reduction. In comparison, the viscosity decreases by more than a factor of 3 over the same range of $h$ in our experiments. Although the analytical solution for suspensions of active ellipsoids appropriate for bacteria is not available, numerical solutions show that the magnitude of $\sigma^S$ of active ellipsoids is smaller than that of active spheres (Takatori and Brady, 2017). Hence, we expect that the effect of the confinement-induced viscosity reduction is even weaker for suspensions of active ellipsoids. This conclusion is supported by the failure of our attempt to collapse data using a dimensionless shear rate with the characteristic run-time of bacteria (Fig. 2b). Thus, other factor(s) need to be considered to fully explain the experimental observations.

**Effect of upstream swimming bacteria at boundary**

A possible origin of $\langle V_d \rangle$ is the boundary layer of upstream swimming bacteria. To estimate their effect, we numerically calculate $\langle V_d \rangle$ by assuming a smectic layer of upstream swimming bacteria in a 2D channel with a line density $n$ randomly distributed along the top and bottom walls (Fig. 7a). Since a bacterium is force-free at low Reynolds numbers, the far-field disturbance flow induced by the bacterium is a hydrodynamic dipolar flow. The flow of a single force dipole near the walls can be constructed by superposition of two Stokeslets of strength $F$ at position $\boldsymbol{r_0} + \boldsymbol{u}d/2$ and $\boldsymbol{r_0} - \boldsymbol{u}d/2$, respectively, where $\boldsymbol{r_0} = (x_0, y_0)$ is the center of the force dipole (Fig. 7b inset). We fix $y_0$ at 1 μm away from the walls in our calculation. The classical solution of a Stokeslet near a wall is used in this construction (Blake 1971), so that the disturbance flow satisfies both the no-penetration and the no-slip boundary conditions at the closest wall. $d$ = 2.2 μm is the distance between the two Stokeslets in a force dipole obtained from experiments (Drescher et al. 2011). $\boldsymbol{u}$ indicates the direction of the force dipole, which forms an acute angle $\theta$ against the imposed shear flow. We choose $\theta = 20°$, the mean angle of bacterial orientation from our experimental observations. Other acute angles lead to qualitatively similar results. $F$ is related to bacterial swimming speed $v$ through $F = \zeta v$ with $F \approx 0.43$ pN for *E. coli* (Drescher et al. 2011). The disturbance flow is translationally invariant along $x$ for infinite large systems. In our calculation, we set the system size to be 1 mm in $x$, much larger than the average spacing between dipoles. The disturbance flow is then obtained near the center of the system. Specifically, we compute the $x$-component disturbance flow, $V_d(y)$, at a fixed $x$ defined as $x = 0$. All the results are averaged over 100 different random configurations. To avoid the singularity at $x = 0$, we exclude those configurations where the locations of the Stokeslets fall in $x \in$ [-10 nm, 10 nm]. A cutoff, $l_c$, in $y$ can be further chosen based on $\int_{-h/2}^{-h/2+l_c} V_d(y)\, dy = 0$, below which the near-field bacterial flow is important and the dipole-flow approximation of our simple model breaks down. $l_c \sim$ 1.4 μm, slightly higher the location of the upper Stokeslet.

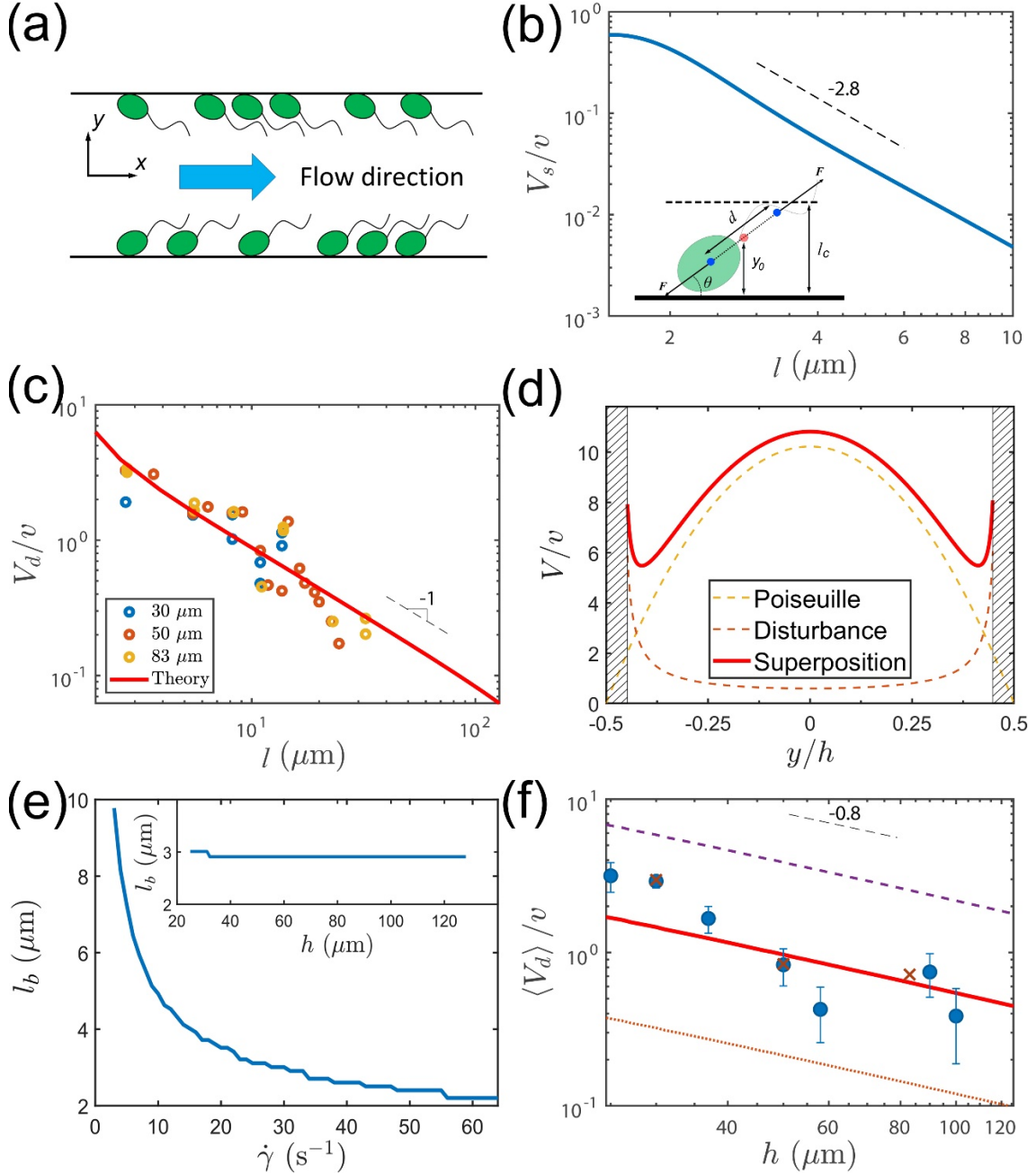

**Fig. 7** Boundary-layer model. (a) Schematic of our numerical model. (b) Velocity induced by a single dipole, $V_s$, as a function of the distance to a flat wall, $l$. The dashed line indicates a power-law scaling $V_s \sim l^{-2.8}$. The inset shows the force dipole of a single bacterium next to the wall with all relevant parameters defined

in the main text. The pink disk indicates the center of the force dipole, whereas the blue disks indicate the locations of the upper and lower Stokeslets. (c) Velocity profile induced by a line of randomly distributed dipoles at dipolar line density $n = 2$ μm$^{-1}$. Symbols are from experiments at different channel heights, which are indicated in the figure. The dashed line indicates a power-law scaling $V_d \sim l^{-1}$. (d) Velocity profiles in a channel of height $h = 30$ μm and shear rate $\dot{\gamma} = 30$ s$^{-1}$. The brown dashed line is the disturbance velocity profile, $V_d(y)$. The yellow dashed line is the parabolic Poiseuille flow. The red solid line is the superposed velocity profile, $V(y)$. The shaded regions are within the cutoff $l_c$. (e) Thickness of the boundary layer, $l_b$, versus $\dot{\gamma}$ with $h = 30$ μm and $n = 2$ μm$^{-1}$. Inset: $l_c$ as function of $h$ with $\dot{\gamma} = 30$ s$^{-1}$. (f) The average disturbance flow velocity $\langle V_d \rangle$ as a function of $h$ at different bacterial densities. Symbols are from experiments in Fig. 6a. Bacterial density: $n = 0.2$ μm$^{-1}$ (brown dotted line), 1.0 μm$^{-1}$ (red solid line) and 4 μm$^{-1}$ (purple dashed line). The black dashed line indicates a power-law scaling $\langle V_d \rangle \sim h^{-0.81}$. All the velocities are normalized by the swimming speed of bacteria $v = 22$ μm/s.

For a single bacterium near wall, the disturbance flow decays as $V_s \sim l^{-2.8}$, faster than the decay of a dipolar flow in bulk, where $l$ is the distance away from wall (Fig. 7b). Collectively, the disturbance flow from the smectic layer of randomly distributed bacteria decays much slower as $V_d \sim l^{-1}$, quantitatively agreeing with our velocity-profile measurements (Fig. 7c). By comparing the experiments and calculation, we obtain the effective dipole density at wall $n = 2$ μm$^{-1}$. Near the center of the channel away from the walls, $V_d$ is small. The fluid is dominated by the pressure-driven Poiseuille flow and is approximately parabolic in shape, consistent with our experiments (Fig. 7d). We estimate the thickness of the boundary layer, $l_b$, by the location where the superposed velocity profile shows a minimum. Below $l_b$, the disturbance flow from the upstreaming bacteria dominates the Poiseuille flow. We find that $l_b$ is a weak function of $h$ over the range of $\dot{\gamma}$ in our experiments (Fig. 7e inset), confirming that the disturbance flow induced by the upstreaming bacteria is confined within a boundary layer. Moreover, $l_b$ decreases with increasing $\dot{\gamma}$ (Fig. 7e). At high $\dot{\gamma}$, the boundary layer becomes insignificant, consistent with our observation (Fig. 2c). At $\dot{\gamma} = 30$ s$^{-1}$, $l$ is about 3 μm, qualitatively agreeing with our observations (see Results).

Finally, the average strength of the disturbance flow, $\langle V_d \rangle$, is calculated by integrating the disturbance velocity profile over the channel height. $\langle V_d \rangle$ as a function of $h$ is shown in Fig. 7f. At a fixed $h$, $\langle V_d \rangle$ increases linearly with bacterial density. More importantly, $\langle V_d \rangle$ decreases with $h$, agreeing with our experimental observations qualitatively (Fig. 7f). The best fitting gives $n = 1.0$ μm$^{-1}$, comparable with the fitting from $V_d(l)$. Numerically, we find $\langle V_d \rangle \sim h^{-0.81}$, with a scaling exponent smaller than that of the experiments. Although the disturbance flow induced by a single bacterium near wall decays as $l^{-2.8}$, too weak to influence the macroscopic fluid flows driven by pressure gradients, the formation of the boundary layer allows such a weak effect from individual bacteria to add coherently, which strongly modifies the fluid flows of bacterial suspensions in confined channels. Hydrodynamic simulations have further shown that coherent structures of active particles including layers of upstream swimming bacteria emerge in confined geometries when the run length of active particles exceeds the confinement length (Nash et al. 2010). Such a finding provides a possible explanation why the confinement effect is most obvious when the system size is comparable or below the run length of bacteria in our experiments.

Although the velocity profiles and the power-law scaling can be qualitatively explained by the presence of the boundary layers of upstream swimming bacteria, there are still open questions on the hypothesis. First, at the experimental bacterial density of ~ 0.2 μm$^{-1}$, the strength of the disturbance flow $V_d$ calculated from our simple model is about 5 times weaker than that from our experiments (Fig. 7f). Such a discrepancy may be accounted for by considering a finite thickness of the boundary layer and including multiple bacterial layers near the confined walls. Second, the calculated velocity profile oscillates strongly within the cutoff length of the boundary layer due to the singular nature of the dipole flow (Fig. 7d). The dipole flow describes the far-field flow of bacteria quantitatively but provides only a qualitative trend in the near field (Drescher et al. 2011; Mathijssen et al. 2016). Thus, to quantify the disturbance flow within the boundary layer close to bacteria, detailed understanding of the near-field bacterial flow is needed. Third, the upstream swimming behavior of bacteria depends on the strength of external flows (Kaya and Koser 2012). Bacterial concentrations next to the walls are also a function of shear rate and system size (Fig. 5b). These effects will likely lead to the dependence of $V_d$ on $\dot{\gamma}$ and $h$, which are not included in our qualitative discussion above. The lack of these dependences may explain the smaller scaling exponent of $\langle V_d \rangle(h)$ in our model. Lastly, it is worth noting that both active hydrodynamic and diffusive stresses are still present in confined bacterial suspensions. In the bulk limit, the contribution of the boundary-driven disturbance flow diminishes. Active hydrodynamic stresslets and diffusive stresses become the leading reason of the non-trivial rheology of active fluids (Alonso-Matilla et al. 2016; Takatori and Brady 2017). Thus, their contributions should also be included in formulating a quantitative theory of active fluids under confinement. We hope our simple heuristic model on the upstream swimming bacteria, an effect that is missing in previous theories, can stimulate further theoretical development.

## Conclusions

We experimentally studied the rheology of confined bacterial suspensions using a microfluidic viscometer. A confinement effect was observed at low shear rates, where the viscosity of bacterial suspensions decreases with increasing confinement. We showed that such a confinement effect arises from the interplay between bacterial motility and confining surfaces. A simple analysis was developed to reveal the physical origin of the confinement effect. We proposed that the boundary layers near confining surfaces, where bacteria collectively swim against the imposed shear flows, play a key role in determining the flow structure and the rheology of confined bacterial suspensions. A simple model based on this picture shows a qualitative agreement with our macroscopic rheology and microscopic dynamics measurements. Open questions were finally discussed for future theoretical development. As such, our experiments demonstrate the importance of system boundary on the flow of bacterial suspensions and present a benchmark to verify different models of the rheology of active fluids. The results are also potentially useful in designing new strategies to modify bacterial transport in confined geometries.

**Acknowledgements** We thank Pranav Agrawal, Shuo Guo, Yi Peng and Yi-Shu Tai for help with experiments and the Dorfman group at UMN for allowing us to use their microfluidics fabrication equipment.

**Funding information** The research was supported by Defense Advanced Research Projects Agency (DARPA), Award Number YFA-D16AP00120 and by National Science Foundation, Award Number CBET-1702352. Portions of this work were conducted in the Minnesota Nano Center, which is supported by the National Science Foundation through the National Nano Coordinated Infrastructure Network, Award Number NNCI-1542202.